\documentclass[table,x11names]{article} 
\usepackage{iclr2025_ai4na, times}

\usepackage{amsmath,amsfonts,bm}

\def\eqref#1{equation~\ref{#1}}

\def\1{\bm{1}}

\DeclareMathAlphabet{\mathsfit}{\encodingdefault}{\sfdefault}{m}{sl}
\SetMathAlphabet{\mathsfit}{bold}{\encodingdefault}{\sfdefault}{bx}{n}

\usepackage{hyperref}
\usepackage{url}
\usepackage{graphicx}

\title{When repeats drive the vocabulary: a Byte-Pair Encoding analysis of T2T primate genomes}

\author{Marina Popova \\
aglabx \\
Paphos, Cyprus \\
\texttt{marinaalexandrovnapopova@gmail.com}
\And
Iaroslav Chelombitko \\
JetBrains \\
Neapolis University Pafos \\
Paphos, Cyprus \\
\texttt{i.chelombitko@nup.ac.cy} \\
\And
Aleksey Komissarov \\
aglabx \\
Paphos, Cyprus \\
\texttt{ad3002@gmail.com} \\
}

\iclrfinalcopy 
\begin{document}

\maketitle

\begin{abstract}
The emergence of telomere-to-telomere (T2T) genome assemblies has opened new avenues for comparative genomics, yet effective tokenization strategies for genomic sequences remain underexplored. In this pilot study, we apply Byte-Pair Encoding (BPE) to nine T2T primate genomes—including three human assemblies—by training independent BPE tokenizers with a fixed vocabulary of 512,000 tokens using our custom tool, dnaBPE. Our analysis reveals that only 11,569 tokens are shared across all assemblies, while nearly 991,854 tokens are unique to a single genome, indicating a rapid decline in shared vocabulary with increasing assembly comparisons. Moreover, phylogenetic trees derived from token overlap failed to recapitulate established primate relationships, a discrepancy attributed to the disproportionate influence of species-specific high-copy repetitive elements. These findings underscore the dual nature of BPE tokenization: while it effectively compresses repetitive sequences, its sensitivity to high-copy elements limits its utility as a universal tool for comparative genomics. We discuss potential hybrid strategies and repeat-masking approaches to refine genomic tokenization, emphasizing the need for domain-specific adaptations in the development of large-scale genomic language models. The dnaBPE tool used in this study is open-source and available at \url{https://github.com/aglabx/dnaBPE}.
\end{abstract}

\section{Introduction}

The recent development of high-quality, telomere-to-telomere (T2T) genome assemblies \citep{nurk2022complete, Liu2024} has greatly enhanced our understanding of genomic structure and evolution among primates \citep{Yoo2025}. Historically, the human reference genome has served as a cornerstone for numerous comparative genomics studies; however, the advent of additional T2T primate assemblies enables more precise examinations of sequence variation, repeat distribution, and lineage-specific genomic features \citep{Yoo2025}. In parallel, the emergence of large language models (LLMs) has ushered in a new era in natural language processing \citep{attention, gpt3, deepseek}, where tokenization methods—most notably Byte-Pair Encoding (BPE, \citep{bpe}), one of the most commonly used tokenizers—have had a profound impact on the field. Although the application of these advanced computational tools to genomic analyses is still in its early stages, their potential to further enhance our understanding of genetic data is actively being explored \citep{benegas2024genomic}. 

Byte-Pair Encoding is designed to efficiently represent repeated substrings by iteratively merging frequent pairs of characters or tokens. While BPE remains a prominent tokenization method in genomic language models, the field has explored several alternative approaches. These include overlapping and non-overlapping k-mer tokenization, as well as nucleotide-level processing. Recent models like HyenaDNA \citep{HyenaDNA}, regLM \citep{Lal2024}, and Caduceus \citep{Caduceus} have opted for nucleotide-level tokenization, while others such as DNABERT \citep{Ji2021} and GeneBERT \citep{GeneBERT} utilize overlapping k-mers. The diversity of approaches reflects the ongoing search for optimal ways to represent genomic sequences in machine learning contexts. However, BPE's widespread adoption in natural language processing has led to its application in genomic analysis, despite fundamental differences between natural language and genetic sequences.

 In NLP tasks, BPE helps reduce large vocabularies by systematically handling rare words and sub-words. However, when applied to eukaryotic genomic sequences—where repeats are not merely noise but biologically significant features such as satellite DNA, transposable elements, and other high-copy regions \citep{Liao2023} — a number of new questions arise. For instance, if BPE is used for tokenization, what impact do these repetitions have on the resulting tokens? Does the prevalence of repeated sequences lead to overly generalized tokens that might obscure subtle, yet crucial, biological variations? Or, conversely, could the inherent repetition in DNA enable a more efficient compression that highlights lineage-specific repeat expansions or contractions? Understanding how repeats influence token formation and distribution is essential for optimizing tokenization strategies in genomic analyses, and may ultimately reveal new insights into the organization and evolution of the genome.

BPE is also used for tokenizing DNA sequences to train models  \citep{benegas2024genomic}. This raised our suspicion that, due to the inherent logic of BPE, it would not be enriched with unique sequences but rather with repetitive elements, which are abundant in the genome. Similar challenges in tokenization strategy selection have also been noted in recent studies \citep{vishniakov2024genomic}, which reported that current pretraining methods for genomic models do not always improve their ability to capture clinically significant genetic variations. Moreover, the genome contains vast regions of satellite DNA sequences that can make up as much as 5\% of the genome and consist of highly similar sequences \citep{Schmid1975}. These regions are likely to be the primary sites from where BPE tokens generated. An intriguing fact is that satellite DNA is species-specific. While it is typically characteristic of groups of species, these groups can still be evolutionarily distant from each other \citep{thakur2021sequence}.

We present a pilot study in which BPE tokenizers of fixed size (512K tokens) were independently trained on nine T2T primate genomes, including three human assemblies (one of which is  the CHM13hTERT human cell line - CHM13). We describe how the vocabularies overlap—or fail to overlap—across these tokenizers, the surprising phylogenetic relationships that emerge when comparing them, and the extent to which BPE could serve as a robust tool for downstream genomics analysis. Ultimately, we evaluate whether BPE tokenization captures meaningful genome-wide features or if it is limited when applied as a general-purpose tokenizer for large-scale model training in genomics. BPE proves to be an inadequate tool for tokenizing multiple genomes in their entirety. The question of which tokenizer is optimal for DNA sequences remains unresolved and, in our opinion, is a primary factor contributing to the challenges in effectively training DNA models.

\section{Materials and Methods}

\subsection{Genome Assemblies}

We selected nine T2T primate genomes, including three distinct human genomes (one of them being the CHM13 reference assembly). This set encompassed both closely and distantly related primates to represent a broad phylogenetic range. All assemblies were curated to ensure they met T2T standards, with minimal gaps and well-resolved centromeric and telomeric regions, including two human genomes HG002 (H\_HG) and Han Chinese (H\_CN)  \citep{jarvis2022semi, yang2023diploid}, and the CHM13hTERT human cell line (H\_CL)  \citep{nurk2022complete}. The non-human primate set comprises representatives of key ape lineages. In our dataset, the Asian great apes are represented by the orangutans, including the Sumatran orangutan (\textit{Pongo abelii}) and the Bornean orangutan (\textit{Pongo pygmaeus}). The small apes are represented by the siamang gibbon (\textit{Symphalangus syndactylus}), which, despite its basal divergence from the great apes, provides essential comparative insights. Among the African apes, we include the chimpanzee (\textit{Pan troglodytes}), bonobo (\textit{Pan paniscus}), and the western lowland gorilla (\textit{Gorilla gorilla}). These species reflect varying degrees of relatedness: while chimpanzees and bonobos are our closest living relatives, gorillas occupy a slightly more basal position within the African ape clade. For every genome except CHM13, we combined the paternal and maternal genomes, after that, we added its reverse complement. As a result, the total genome size amounted to approximately 12 billion base pairs of DNA and 6 billion base pairs for CHM13.

\subsection{DNA-specific BPE tokenizer}

\begin{table}[t]
\vspace{1em} 
\caption{First 42 tokens from BPE tokenization (vocabulary size 512,000) across nine T2T primate genomes, showing perfect conservation of initial dinucleotides, followed by increasing divergence patterns that reflect both evolutionary relationships and species-specific repetitive elements. Color coding (green for conservation, orange for divergence) reveals three distinct patterns: high conservation in human assemblies through rank 37, intermediate conservation within great apes, and pronounced species-specific variations in more distant primates.}
\label{tab:primate-seq}
\vspace{1em} 
\begin{center}
\small
\setlength{\tabcolsep}{5pt}  
\renewcommand{\arraystretch}{1.1} 
\begin{tabular}{c c c c c c c c c c}
\hline
\textbf{Rank} & \textbf{H\_HG} & \textbf{H\_CN} & \textbf{H\_CL} & \textbf{Sum\_orang} & \textbf{Gibbon} & \textbf{Chimp} & \textbf{Born\_orang} & \textbf{Gorilla} & \textbf{Bonobo} \\
\hline
1  & A    & A    & A    & A    & A    & A    & A    & A    & A \\
2  & C    & C    & C    & C    & C    & C    & C    & C    & C \\
3  & G    & G    & G    & G    & G    & G    & G    & G    & G \\
4  & T    & T    & T    & T    & T    & T    & T    & T    & T \\
\rowcolor{green!10}
\cellcolor{white} 5  & AA   & AA   & AA   & AA   & AA   & AA   & AA   & AA   & AA \\
\rowcolor{green!10}
\cellcolor{white} 6  & TT   & TT   & TT   & TT   & TT   & TT   & TT   & TT   & TT \\
\rowcolor{green!10}
\cellcolor{white} 7  & TG   & TG   & TG   & TG   & TG   & TG   & TG   & TG   & TG \\
8  & \cellcolor{green!10} AG   & \cellcolor{green!10} AG   & \cellcolor{green!10} AG   & \cellcolor{green!10} AG   & \cellcolor{green!10} AG   & \cellcolor{green!10} AG   & \cellcolor{orange!15}CA & \cellcolor{orange!15}CA & \cellcolor{orange!15}CA \\
9  & \cellcolor{green!10} CC   & \cellcolor{green!10} CC   & \cellcolor{green!10} CC   & \cellcolor{green!10} CC   & \cellcolor{green!10} CC   & \cellcolor{green!10} CC   & \cellcolor{green!10} CC   & \cellcolor{orange!15}TA & \cellcolor{orange!15}TA \\
10 & \cellcolor{green!10} TC   & \cellcolor{green!10} TC   & \cellcolor{green!10} TC   & \cellcolor{green!10} TC   & \cellcolor{green!10} TC   & \cellcolor{green!10} TC   & GG   & \cellcolor{orange!15} CC   & \cellcolor{orange!15} CC \\
11 & \cellcolor{green!10} AC   & \cellcolor{green!10} AC   & \cellcolor{green!10} AC   & \cellcolor{green!10} AC   & \cellcolor{green!10} AC   & \cellcolor{green!10} AC   & TA   & \cellcolor{orange!15} GG   & \cellcolor{orange!15} GG \\
12 & \cellcolor{green!10} GG   & \cellcolor{green!10} GG   & \cellcolor{green!10} GG   & \cellcolor{green!10} GG   & \cellcolor{green!10} GG   & \cellcolor{green!10} GG   & TC   & \cellcolor{orange!15} TC   & \cellcolor{orange!15} TC \\
13 & \cellcolor{green!10} ATT  & \cellcolor{green!10} ATT  & \cellcolor{green!10} ATT  & \cellcolor{green!10} ATT  & \cellcolor{green!10} ATT  & AT   & GA   & \cellcolor{orange!15} GA   & \cellcolor{orange!15} GA \\
14 & \cellcolor{green!10} AT   & \cellcolor{green!10} AT   & \cellcolor{green!10} AT   & \cellcolor{green!10} AT   & \cellcolor{green!10} AT   & ATT  & AAA  & \cellcolor{orange!15} AAA  & \cellcolor{orange!15} AAA \\
15 & \cellcolor{green!10} ATG  & \cellcolor{green!10} ATG  & \cellcolor{green!10} ATG  & \cellcolor{green!10} ATG  & \cellcolor{green!10} ATG  & ATG  & GC   & \cellcolor{orange!15} GC   & \cellcolor{orange!15} GC \\
16 & \cellcolor{green!10} GC   & \cellcolor{green!10} GC   & \cellcolor{green!10} GC   & \cellcolor{green!10} GC   & \cellcolor{green!10} GC   & GC   & TAA  & \cellcolor{orange!15} TAA  & \cellcolor{orange!15} TAA \\
17 & \cellcolor{green!10} TAA  & \cellcolor{green!10} TAA  & \cellcolor{green!10} TAA  & \cellcolor{green!10} TAA  & \cellcolor{green!10} TAA  & TAA  & TCA  & GAA  & TCA \\
18 & \cellcolor{green!10} TCC  & \cellcolor{green!10} TCC  & \cellcolor{green!10} TCC  & \cellcolor{green!10} TCC  & AAAA & TCC  & TGA  & TTC  & TGA \\
19 & \cellcolor{green!10} ACC  & \cellcolor{green!10} ACC  & \cellcolor{green!10} ACC  & AAAA & TCC  & ACC  & TTTT & TCA  & TTTT \\
20 & \cellcolor{green!10} AAAA & \cellcolor{green!10} AAAA & \cellcolor{green!10} AAAA & ACC  & ACC  & ATC  & GAA  & TGA  & GAA \\
21 & \cellcolor{green!10} AGG  & \cellcolor{green!10} AGG  & \cellcolor{green!10} AGG  & AGG  & TTC  & AAAA & TCC  & TTA  & TCC \\
22 & \cellcolor{green!10} ATC  & \cellcolor{green!10} ATC  & \cellcolor{green!10} ATC  & TTC  & AGC  & AGG  & TTC  & TCC  & TTA \\
23 & \cellcolor{green!10} TTC  & \cellcolor{green!10} TTC  & \cellcolor{green!10} TTC  & AGC  & AGG  & AGC  & TTA  & TTTT & TTC \\
24 & \cellcolor{green!10} AGC  & \cellcolor{green!10} AGC  & \cellcolor{green!10} AGC  & ATC  & ATC  & TTC  & GTG  & GTG  & GTG \\
25 & \cellcolor{green!10} TGG  & \cellcolor{green!10} TGG  & \cellcolor{green!10} TGG  & TGG  & TGC  & TGG  & CAC  & CAC  & CTG \\
26 & \cellcolor{green!10} AAG  & \cellcolor{green!10} AAG  & \cellcolor{green!10} AAG  & AAG  & TTTT & AAG  & AAAA & AAAA & AAAA \\
27 & \cellcolor{green!10} TGC  & \cellcolor{green!10} TGC  & \cellcolor{green!10} TGC  & TGC  & AAG  & TGC  & GGA  & GGA  & GGA \\
28 & \cellcolor{green!10} TTTT & \cellcolor{green!10} TTTT & \cellcolor{green!10} TTTT & TTTT & AAC  & TTTT & CTG  & GTT  & GCA \\
29 & \cellcolor{green!10} AAC  & \cellcolor{green!10} AAC  & \cellcolor{green!10} AAC  & AAC  & TGG  & AAC  & GCA  & CTG  & CCA \\
30 & \cellcolor{green!10} TTG  & \cellcolor{green!10} TTG  & \cellcolor{green!10} TTG  & TTG  & TAG  & TAG  & GCC  & CATG & GCC \\
31 & \cellcolor{green!10} TAG  & \cellcolor{green!10} TAG  & \cellcolor{green!10} TAG  & TAG  & TAC  & TAC  & GTT  & GCC  & CAA \\
32 & \cellcolor{green!10} TAC  & \cellcolor{green!10} TAC  & \cellcolor{green!10} TAC  & TAC  & TTG  & TTG  & CAA  & TCTG & GTT \\
33 & \cellcolor{green!10} CCC  & \cellcolor{green!10} CCC  & \cellcolor{green!10} CCC  & CCC  & AGAA & TAT  & GTA  & TATT & CTT \\
34 & \cellcolor{green!10} TATT & \cellcolor{green!10} TATT & \cellcolor{green!10} TATT & TATT & TATT & CCC  & CTT  & GCA  & TAAA \\
35 & \cellcolor{green!10} TGGG & \cellcolor{green!10} TGGG & \cellcolor{green!10} TGGG & TGGG & CCC  & TATT & CCA  & TAC  & TATA \\
36 & \cellcolor{green!10} TAT  & \cellcolor{green!10} TAT  & \cellcolor{green!10} TAT  & AGAA & TTTC & TGGG & TCTG & TAAA & TCTG \\
37 & \cellcolor{green!10} AGAA & \cellcolor{green!10} AGAA & \cellcolor{green!10} AGAA & TAT  & TAT  & ACAG & TAAA & GAAA & GTA \\
38 & \cellcolor{green!10} TTTC & \cellcolor{green!10} TTTC & ATTC & TTTC & TGGG & AAAG & GAAA & TACA & TCCA \\
39 & \cellcolor{green!10} AGGG & \cellcolor{green!10} AGGG & TTTC & AGGG & ATTC & AAAC & TGTG & CAA  & GAAA \\
40 & \cellcolor{green!10} ATTC & \cellcolor{green!10} ATTC & AGGG & TGTG & AGAG & ATGG & TCTT & CTT  & TATT \\
41 & \cellcolor{green!10} AGGC & \cellcolor{green!10} AGGC & AGGC & ATTC & AAAC & AAGG & CTA  & GTA  & TCTT \\
42 & \cellcolor{green!10} TGTG & \cellcolor{green!10} TGTG & TGTG & AGGC & TGTG & TTTC & CACA & TCTT & CTA \\
\hline
\end{tabular}
\end{center}
\vspace{1em}
\end{table}

For such a large genome size, existing tools are unable to efficiently tokenize the data without an enormous cost in memory or time. For example, the Hugging Face tokenizer simply freezes when processing such a volume of data. Another issue is that conventional tokenizers are designed for natural language text and contain a large number of unnecessary tokens. 

Therefore, we decided to implement dnaBPE tokenizer specifically optimized for tokenizing large-scale DNA sequences: dnaBPE. We introduced several novel features to our tokenizer that are uncommon in text-based tokenizers. First, we processed only the four nucleotide bases—A, T, G, and C—while treating all other characters as the beginning of a new sequence. The next innovation was the addition of two extra containers that proved useful for post-analysis. The first container stores the frequency of a pair at the moment of its merging. The second records the genomic positions of merged pairs, storing a list of the sequence ID and the position within that sequence. This significantly simplified our subsequent data analysis.

\subsection{Tokenizer Construction and Token Sets}

For each of the nine genomes, we independently trained a BPE tokenizer of 512,000 tokens. This size was selected to capture the deep diversity of possible k-mers while also reflecting the typical upper range used in large language models. The current implementation is written in C++ and is available in our GitHub repository. It is currently single-threaded, but we plan to make it multithreaded in the future, which will significantly improve processing speed. In its current form, running on a single Intel(R) Xeon(R) Platinum 8268 CPU @ 2.90GHz, tokenizing a 12 GB genome takes approximately 5 hours and requires 200 GB of RAM.

After obtaining the vocabularies, we compared and quantified overlaps between the nine token sets. We computed how many tokens were shared by exactly 1, 2, 3, up to all 9 tokenizers. These shared (or unique) tokens were tabulated to reveal the degree of overlap. Additionally, we refer to the combined dataset of common tokens from nine assemblies as the core token dataset.

\subsection{Phylogenetic Analysis of Token Overlaps}

To explore whether BPE token overlap could recapitulate known evolutionary relationships, we calculated pairwise distances between each pair of tokenizers. These distances were derived based on the similarity or dissimilarity of token sets and their frequencies. We then constructed a phylogenetic-like tree from these distances to assess whether the clustering of genomes followed expected primate relationships.

\subsection{Token annotation}

For token annotation, we used the diploid human genome T2T HG002.v.1.1, specifically its maternal haplotype (HG002.v.1.1.mat.fasta \url{https://github.com/marbl/HG002}). For this genome, we obtained protein coding exons annotations derived via liftover from \url{s3://human-pangenomics/T2T/HG002/assemblies/annotation/JHULiftoff/Jul2024/HG002v110.JHU.20240718.bb}.

Additionally, we incorporated annotations from RepeatMasker \url{s3://human-pangenomics/T2T/HG002/assemblies/annotation/repeatmasker/hg002v1.1.maternal.bb} and cenSat v2.0 \url{s3://human-pangenomics/T2T/HG002/assemblies/annotation/centromere/hg002v1.1_v2.0/hg002v1.1.cenSatv2.0.MAT.bed}.

From these annotations, we extracted Human Satellite and Alpha Satellite sequences, and among dispersed repeats, we identified the positions of ALU repeats. For coding sequences, we extracted the positions of protein coding exons. For token positions, we utilized the additional fields we incorporated into the tokenizer, which store information linking each token to its corresponding position in the genome.

\subsection{Comparison with Existing BPE Tokenizers from DNA Language Models}

To compare the token sets of other models, we downloaded tokenizers from publicly available popular DNA models that use BPE tokenization. From these tokenizers, we extracted the vocabularies and applied the same analysis that we previously conducted for our core tokens.

Since existing BPE tokenizers were not designed for genome-scale tokenization, we developed a custom C++ tool, \texttt{TokPos}\footnote{\url{https://github.com/aglabx/TokPos}}, to efficiently search for exact matches between tokens and positions in the genome. This program outputs data in the same format as our tokenizer, ensuring a direct and fair comparison between different models.

\begin{figure}[h]
    \centering
    \includegraphics[width=1.0\linewidth]{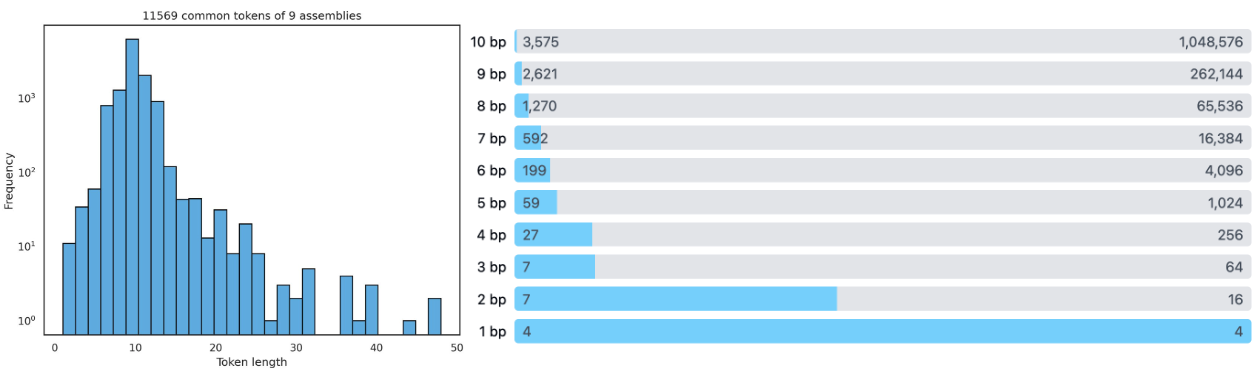}
    \caption{Analysis of token length distributions in the core set of 11,569 tokens shared across nine primate genomes. Left: Frequency histogram (log scale) showing the distribution of token lengths, with a pronounced peak at 8-12 bp and declining frequency of longer sequences. Right: Comparison of observed tokens (blue bars) versus theoretical maximum possible sequences (gray bars) for each length, demonstrating complete representation of 1-bp tokens (4/4), substantial coverage of 2-bp tokens (7/16), and rapidly decreasing coverage for longer sequences (e.g., only 0.34\% of possible 10-bp sequences). This pattern suggests that BPE tokenization effectively captures short, conserved sequence motifs while longer tokens become increasingly species-specific.}
    \label{fig:common_tokens}
\end{figure}

\section{Results}

\subsection{Early BPE steps in nine genomes}
Detailed analysis of the first 42 tokens revealed distinct patterns of conservation and divergence across primate genomes. The initial dinucleotides showed remarkable conservation, with AA, TT, and TG appearing in ranks 5-7 across all nine genomes.
Conservation patterns began to diverge at rank 8, revealing three distinct groupings:

Human assemblies (HG, CN, CL) demonstrated exceptional conservation through rank 37, sharing nearly identical token sequences and order.
Great apes showed intermediate conservation, with Sumatran orangutan sharing patterns with humans through rank 17, while gorilla and Bornean orangutan displayed distinct yet related patterns.
More distant primates, particularly the gibbon and bonobo, showed earlier divergence in their token patterns, often featuring unique sequences not prominently ranked in other species.

The emergence of longer sequences (3-4 bp) showed increasing species specificity, with tokens like AAAA, TTTT, and TGGG appearing at different ranks across species. This pattern suggests that while basic genomic building blocks are conserved, the frequency and organization of longer sequences reflect both evolutionary relationships and species-specific repetitive elements (Table \ref{tab:primate-seq}).

\subsection{Length Distribution of Common Tokens}

Analysis of the 11,569 tokens shared across all nine primate genomes revealed distinctive patterns in sequence length distribution. The most striking feature was the concentration of token lengths between 8-12 base pairs, with a pronounced peak around 10 bp. This distribution pattern offers insights into the nature of conserved genomic elements captured by BPE tokenization.

When comparing observed tokens to theoretical maxima for each length, we found a clear trend of decreasing coverage as sequence length increased. Dinucleotides maintained substantial coverage (7 out of 16 possible combinations). However, this coverage dropped dramatically for longer sequences: only 592 out of 16,384 possible 7-bp sequences (3.6\%) and merely 3,575 out of 1,048,576 possible 10-bp sequences (0.34\%) were shared across all genomes.
This exponential decline in coverage with increasing length suggests two key aspects of genomic sequence conservation: (1) shorter motifs are more likely to be functionally constrained and thus conserved across species, and (2) longer sequences tend to be more species-specific, possibly due to their role in regulatory elements or repetitive structures. The peak at 8-12 bp might represent an optimal length for functional genomic elements that are conserved enough to be captured by BPE tokenization across diverse primate species (Figure \ref{fig:common_tokens}).

\begin{figure}[h]
    \centering
    \includegraphics[width=1.0\linewidth]{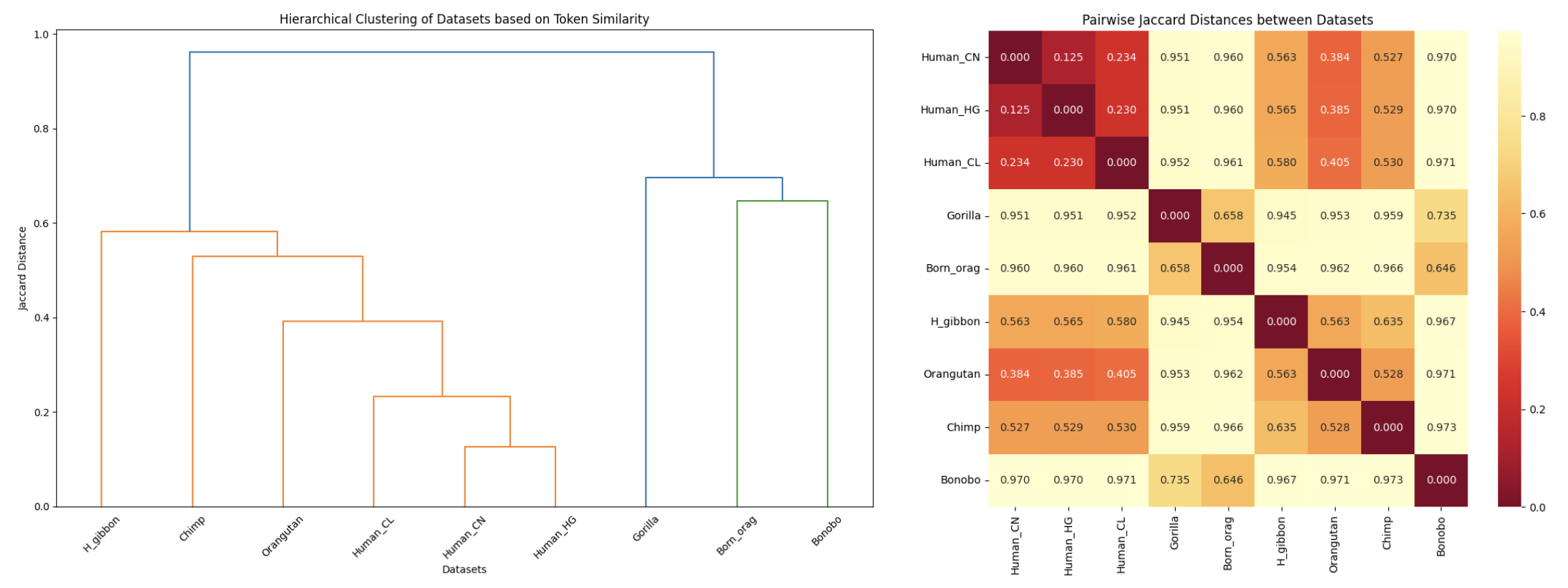}
    \caption{Comparative analysis of BPE token distributions across nine primate genomes. Left: Hierarchical clustering dendrogram based on token similarity reveals unexpected groupings that do not align with known primate phylogeny, with human genomes (HG, CN, CL) scattered across different clusters rather than forming a monophyletic group. Right: Heatmap of pairwise Jaccard distances between datasets shows high similarity among human assemblies (0.125-0.234) but unexpectedly high distances between evolutionarily close species, suggesting that BPE tokenization is strongly influenced by species-specific repetitive elements rather than evolutionary relationships. Color scale ranges from dark red (low distance, high similarity) to pale yellow (high distance, low similarity), highlighting the disconnect between token-based clustering and established primate phylogeny.}
    \label{fig:heatmap}
\end{figure}

\subsection{BPE Token Overlap}
The key question we aimed to answer was how many common tokens exist across the nine tokenizers. Surprisingly, the overlap turned out to be very small: only 11,569 tokens (0.6\% of total tokens) were shared across all nine genomes. Our analysis revealed a striking gradient of token sharing: 991,854 tokens (51.2\%) were unique to individual genomes, while 276,808 (14.3\%) and 231,412 (11.9\%) tokens were shared between two and three genomes, respectively. A notable decline occurred at four genomes, with only 75,874 tokens (3.9\%) shared, followed by a temporary increase for five and six genomes (110,156 and 205,776 tokens), before dropping sharply for higher numbers (Figure \ref{fig:supplementary1}).
This distribution pattern, with such a small number of universally shared tokens and nearly a million genome-specific tokens, suggests that BPE tokenization is strongly influenced by species-specific sequences. The rapid decrease in shared vocabulary as more assemblies are considered highlights fundamental challenges in developing a universal genomic tokenizer, particularly given the prevalence of unique repetitive elements in each genome.

\subsection{Phylogenetic Incongruence}

We performed hierarchical clustering analysis of token distributions to assess whether BPE tokenization could recapitulate known evolutionary relationships among primates. The results revealed surprising patterns that deviated significantly from established primate phylogeny.

The pairwise Jaccard distances analysis showed remarkably high similarity among human assemblies (distances ranging from 0.125 to 0.234), as expected for within-species comparisons. However, the distances between human and other great apes were unexpectedly large, with most values exceeding 0.9, even for evolutionarily close relatives like chimpanzees. The Bornean orangutan and gorilla showed moderate similarity to each other (distance 0.658) but exhibited high distances to other primates, including their close relatives.

The hierarchical clustering dendrogram further emphasized this phylogenetic incongruence. Instead of clustering according to known evolutionary relationships, the genomes formed unexpected groupings. Most strikingly, human genomes (H\_HG, H\_CN, H\_CL) did not form a monophyletic group as expected, but were scattered across different clusters. This pattern suggests that BPE tokenization captures species-specific genomic features, particularly repetitive elements, rather than evolutionary relationships. The results highlight how satellite DNA and other repetitive sequences, which can evolve rapidly and independently, dominate the token distribution patterns over more conserved genomic regions that typically inform phylogenetic relationships (Figure \ref{fig:heatmap}).

\begin{table}[t]
\vspace{1em} 
\caption{Comparison of token characteristics and genomic coverage across different tokenization approaches and expected genome content. Size indicates total vocabulary size; median and maximum token lengths are in base pairs. Coverage percentages are shown for different genomic elements: all repetitive elements (All reps), exons, satellite DNA (SatDNA), Alu elements, and other sequences.
}
\label{tab:dna-features}
\vspace{1em} 
\begin{center}
\small
\begin{tabular}{l c c c c c c c c}
\hline
\textbf{Model} & \textbf{Size} & \textbf{Median tok len} & \textbf{Max tok len} & \textbf{All reps} & \textbf{Exons} & \textbf{SatDNA} & \textbf{Alu} & \textbf{Other} \\
\hline
dnaBPE-9     & 11569 & 10 & 48 & 77.6 & 10.4 & 0.4 & 0.4 & 11.2 \\
GROVER    & 610   & 5  & 16 & 77.8 & 10.5 & 0.2 & 0.4 & 11.1 \\
GENA-LM   & 32000 & 9  & 64 & 77.3 & 10.4 & 0.7 & 0.4 & 11.2 \\
DNABERT-2 & 4096  & 7  & 32 & 77.6 & 10.5 & 0.3 & 0.4 & 11.2 \\
\hline
\end{tabular}
\end{center}
\vspace{1em} 
\end{table}

\subsection{Tokens annotation}

Comparison of our core tokenizer of nine primates with existing approaches (GROVER, GENA-LM, and DNABERT-2) revealed unexpected patterns in genomic element representation. All tokenizers showed similar coverage of exonic regions (approximately 10.4-10.5\%), closely matching biological expectations. However, the representation of repetitive elements deviated significantly from expected genomic content.
Most strikingly, all tokenizers substantially overrepresented repetitive elements, capturing approximately 77-78\% of their vocabulary compared to the expected 54\% genomic content \citep{nurk2022complete}. Conversely, Alu elements were severely underrepresented (0.4\% versus expected 11\% according to published copy number multiplied on element length \citep{Deininger2011}). Satellite DNA showed variable representation across tokenizers (0.2-0.7\%) but remained well below its expected genomic content of 5\%.
These discrepancies suggest that current tokenization approaches, regardless of vocabulary size or token length parameters, share similar biases in their representation of genomic elements. The core tokens from nine primates, despite having a larger vocabulary (11,569 tokens) and longer maximum token length (48 bp), showed similar coverage patterns to smaller models like GROVER (610 tokens). This consistent pattern across different architectures suggests a fundamental limitation in how BPE-based tokenization captures genomic sequence characteristics (Table \ref{tab:dna-features}). 

It is important to note that all tokenizers were trained on similar datasets. Since the human genome is present in all tokenizers, the similarity in token distributions can be explained by the fact that they were trained using the same algorithm on the same data. The shorter the token, the more frequently it appears in the genome, covering a wider range of contexts, and thus exerting a greater influence on the final model. In contrast, longer tokens occur in fewer contexts, making their overall impact significantly smaller. We suggest that increasing the tokenizer size does not significantly improve performance in the case of BPE with DNA data.

\section{Discussion}

Our study demonstrates that BPE tokenization, while capable of capturing high-copy repeats across T2T primate genomes, has significant limitations as a universal tokenizer for comparative genomics. The strikingly low number of shared tokens (0.6\%) and the incongruence between token-based and established primate phylogenies highlight how species-specific repetitive sequences dominate the BPE vocabulary.

Our method offers several practical advantages: it enables tokenizer evaluation before costly model training, assesses generalization potential to new genomes, and suggests ways to optimize tokenizer size by identifying and removing uninformative tokens. For applications focused on identifying and analyzing high-frequency repeats, BPE can serve as a powerful, automated strategy for tasks such as satellite DNA annotation or novel repeat family discovery.
However, for broader comparative genomics applications or training large-scale models, BPE's sensitivity to repetitive elements poses significant challenges. The heavy influence of repeats can overshadow crucial evolutionary signals in protein-coding and regulatory regions, potentially limiting the tokenizer's ability to capture functionally important sequences shared across species. This bias could impair downstream models' capacity to learn cross-species genomic features relevant to gene function, regulation, or disease.

Based on our findings, we propose several practical directions for improving genomic tokenization. A promising approach would be to develop a two-stage tokenization process that handles different genomic elements separately. Known functional elements such as exons, regulatory regions, and conserved non-coding sequences could be tokenized with parameters optimized for their specific characteristics, while repetitive regions would undergo separate processing. This separation would help maintain the biological significance of functional elements while still effectively compressing repetitive sequences.

To address the overwhelming influence of repetitive DNA, we suggest implementing targeted repeat masking before tokenization. Using established tools like RepeatMasker, satellite DNA and other repeats could be selectively masked while preserving other repetitive elements that may have functional significance. This balanced approach would help prevent repeated sequences from dominating the token vocabulary while retaining important biological information.

The tokenization process could be further improved through an adaptive vocabulary system. This would begin with a core set of tokens derived from highly conserved regions, then dynamically expand to include species-specific tokens as needed. The vocabulary size could be adjusted based on sequence conservation levels, allowing for more efficient representation of both shared and unique genomic features.

Our analysis revealed that BPE tokenization tends to be dominated by high-copy repeats, which makes it challenging to identify evolutionarily and functionally significant signals. These findings align with the conclusions of \citep{vishniakov2024genomic}, who observed that even when using genome foundation models, pretrained models often fail to outperform randomly initialized models in tasks requiring high sensitivity to mutations.

This highlights the need to reconsider current approaches to genomic sequence tokenization.

Finally, we see potential in developing a pipeline that leverages shared BPE tokens for comparative genomic analysis. By analyzing token patterns across related species, we could identify and annotate novel repetitive elements, particularly valuable for newly sequenced genomes where traditional annotation resources are limited. These tokens could serve as seeds for discovering new repeat families and understanding genome evolution.

By incorporating biological knowledge into tokenization strategies, we may better harness BPE's compression advantages while mitigating its bias toward repetitive elements. While BPE-derived "token trees" may not reliably reconstruct evolutionary histories, they could provide valuable insights into repeat landscape evolution across lineages.

\section{Data Availability and reproducibility}

Reproducible code is available at \href{https://github.com/aglabx/dnaBPE}{github.com/aglabx/dnaBPE} under the MIT license, and the extended datasets are hosted at \href{https://huggingface.co/datasets/aglabx/primates_BPE}{huggingface.co/datasets/aglabx/primates\_BPE}.


\section{AI models usage}

For text editing, since it was written by non-native English speakers, we used ChatGPT-4o and o3-mini models. Additionally, we utilized Claude Sonnet 3.5 from Anthropic for further editing. However, we did not use any models for text generation. Additionally, we used GitHub Copilot to correct errors in the code.

\bibliography{iclr2025_ai4na}

\begin{thebibliography}{20}
\providecommand{\natexlab}[1]{#1}
\providecommand{\url}[1]{\texttt{#1}}
\expandafter\ifx\csname urlstyle\endcsname\relax
  \providecommand{\doi}[1]{doi: #1}\else
  \providecommand{\doi}{doi: \begingroup \urlstyle{rm}\Url}\fi

\bibitem[Benegas et~al.(2024)]{benegas2024genomic}
Gonzalo Benegas et~al.
\newblock Genomic language models: Opportunities and challenges, 2024.
\newblock URL \url{https://doi.org/10.48550/arXiv.2407.11435}.
\newblock arXiv preprint.

\bibitem[Bostrom \& Durrett(2020)Bostrom and Durrett]{bpe}
Kaj Bostrom and Greg Durrett.
\newblock Byte pair encoding is suboptimal for language model pretraining, 2020.
\newblock URL \url{https://arxiv.org/abs/2004.03720}.

\bibitem[Brown et~al.(2020)Brown, Mann, Ryder, Subbiah, Kaplan, Dhariwal, Neelakantan, Shyam, Sastry, Askell, Agarwal, Herbert-Voss, Krueger, Henighan, Child, Ramesh, Ziegler, Wu, Winter, Hesse, Chen, Sigler, Litwin, Gray, Chess, Clark, Berner, McCandlish, Radford, Sutskever, and Amodei]{gpt3}
Tom~B. Brown, Benjamin Mann, Nick Ryder, Melanie Subbiah, Jared Kaplan, Prafulla Dhariwal, Arvind Neelakantan, Pranav Shyam, Girish Sastry, Amanda Askell, Sandhini Agarwal, Ariel Herbert-Voss, Gretchen Krueger, Tom Henighan, Rewon Child, Aditya Ramesh, Daniel~M. Ziegler, Jeffrey Wu, Clemens Winter, Christopher Hesse, Mark Chen, Eric Sigler, Mateusz Litwin, Scott Gray, Benjamin Chess, Jack Clark, Christopher Berner, Sam McCandlish, Alec Radford, Ilya Sutskever, and Dario Amodei.
\newblock Language models are few-shot learners, 2020.
\newblock URL \url{https://arxiv.org/abs/2005.14165}.

\bibitem[{DeepSeek-AI} et~al.(2025){DeepSeek-AI}, Guo, Yang, Zhang, Song, Zhang, Xu, Zhu, Ma, Wang, Bi, Zhang, Yu, Wu, Wu, Gou, Shao, Li, Gao, Liu, Xue, Wang, Wu, Feng, Lu, Zhao, Deng, Zhang, Ruan, Dai, Chen, Ji, Li, Lin, Dai, Luo, Hao, Chen, Li, Zhang, Bao, Xu, Wang, Ding, Xin, Gao, Qu, Li, Guo, Li, Wang, Chen, Yuan, Qiu, Li, Cai, Ni, Liang, Chen, Dong, Hu, Gao, Guan, Huang, Yu, Wang, Zhang, Zhao, Wang, Zhang, Xu, Xia, Zhang, Zhang, Tang, Li, Wang, Li, Tian, Huang, Zhang, Wang, Chen, Du, Ge, Zhang, Pan, Wang, Chen, Jin, Chen, Lu, Zhou, Chen, Ye, Wang, Yu, Zhou, Pan, Li, Zhou, Wu, Ye, Yun, Pei, Sun, Wang, Zeng, Zhao, Liu, Liang, Gao, Yu, Zhang, Xiao, An, Liu, Wang, Chen, Nie, Cheng, Liu, Xie, Liu, Yang, Li, Su, Lin, Li, Jin, Shen, Chen, Sun, Wang, Song, Zhou, Wang, Shan, Li, Wang, Wei, Zhang, Xu, Li, Zhao, Sun, Wang, Yu, Zhang, Shi, Xiong, He, Piao, Wang, Tan, Ma, Liu, Guo, Ou, Wang, Gong, Zou, He, Xiong, Luo, You, Liu, Zhou, Zhu, Xu, Huang, Li, Zheng, Zhu, Ma, Tang, Zha, Yan, Ren, Ren, Sha, Fu, Xu, Xie,
  Zhang, Hao, Ma, Yan, Wu, Gu, Zhu, Liu, Li, Xie, Song, Pan, Huang, Xu, Zhang, and Zhang]{deepseek}
{DeepSeek-AI}, Daya Guo, Dejian Yang, Haowei Zhang, Junxiao Song, Ruoyu Zhang, Runxin Xu, Qihao Zhu, Shirong Ma, Peiyi Wang, Xiao Bi, Xiaokang Zhang, Xingkai Yu, Yu~Wu, Z.~F. Wu, Zhibin Gou, Zhihong Shao, Zhuoshu Li, Ziyi Gao, Aixin Liu, Bing Xue, Bingxuan Wang, Bochao Wu, Bei Feng, Chengda Lu, Chenggang Zhao, Chengqi Deng, Chenyu Zhang, Chong Ruan, Damai Dai, Deli Chen, Dongjie Ji, Erhang Li, Fangyun Lin, Fucong Dai, Fuli Luo, Guangbo Hao, Guanting Chen, Guowei Li, H.~Zhang, Han Bao, Hanwei Xu, Haocheng Wang, Honghui Ding, Huajian Xin, Huazuo Gao, Hui Qu, Hui Li, Jianzhong Guo, Jiashi Li, Jiawei Wang, Jingchang Chen, Jingyang Yuan, Junjie Qiu, Junlong Li, J.~L. Cai, Jiaqi Ni, Jian Liang, Jin Chen, Kai Dong, Kai Hu, Kaige Gao, Kang Guan, Kexin Huang, Kuai Yu, Lean Wang, Lecong Zhang, Liang Zhao, Litong Wang, Liyue Zhang, Lei Xu, Leyi Xia, Mingchuan Zhang, Minghua Zhang, Minghui Tang, Meng Li, Miaojun Wang, Mingming Li, Ning Tian, Panpan Huang, Peng Zhang, Qiancheng Wang, Qinyu Chen, Qiushi Du, Ruiqi Ge,
  Ruisong Zhang, Ruizhe Pan, Runji Wang, R.~J. Chen, R.~L. Jin, Ruyi Chen, Shanghao Lu, Shangyan Zhou, Shanhuang Chen, Shengfeng Ye, Shiyu Wang, Shuiping Yu, Shunfeng Zhou, Shuting Pan, S.~S. Li, Shuang Zhou, Shaoqing Wu, Shengfeng Ye, Tao Yun, Tian Pei, Tianyu Sun, T.~Wang, Wangding Zeng, Wanjia Zhao, Wen Liu, Wenfeng Liang, Wenjun Gao, Wenqin Yu, Wentao Zhang, W.~L. Xiao, Wei An, Xiaodong Liu, Xiaohan Wang, Xiaokang Chen, Xiaotao Nie, Xin Cheng, Xin Liu, Xin Xie, Xingchao Liu, Xinyu Yang, Xinyuan Li, Xuecheng Su, Xuheng Lin, X.~Q. Li, Xiangyue Jin, Xiaojin Shen, Xiaosha Chen, Xiaowen Sun, Xiaoxiang Wang, Xinnan Song, Xinyi Zhou, Xianzu Wang, Xinxia Shan, Y.~K. Li, Y.~Q. Wang, Y.~X. Wei, Yang Zhang, Yanhong Xu, Yao Li, Yao Zhao, Yaofeng Sun, Yaohui Wang, Yi~Yu, Yichao Zhang, Yifan Shi, Yiliang Xiong, Ying He, Yishi Piao, Yisong Wang, Yixuan Tan, Yiyang Ma, Yiyuan Liu, Yongqiang Guo, Yuan Ou, Yuduan Wang, Yue Gong, Yuheng Zou, Yujia He, Yunfan Xiong, Yuxiang Luo, Yuxiang You, Yuxuan Liu, Yuyang Zhou, Y.~X.
  Zhu, Yanhong Xu, Yanping Huang, Yaohui Li, Yi~Zheng, Yuchen Zhu, Yunxian Ma, Ying Tang, Yukun Zha, Yuting Yan, Z.~Z. Ren, Zehui Ren, Zhangli Sha, Zhe Fu, Zhean Xu, Zhenda Xie, Zhengyan Zhang, Zhewen Hao, Zhicheng Ma, Zhigang Yan, Zhiyu Wu, Zihui Gu, Zijia Zhu, Zijun Liu, Zilin Li, Ziwei Xie, Ziyang Song, Zizheng Pan, Zhen Huang, Zhipeng Xu, Zhongyu Zhang, and Zhen Zhang.
\newblock Deepseek-r1: Incentivizing reasoning capability in llms via reinforcement learning, 2025.
\newblock URL \url{https://arxiv.org/abs/2501.12948}.

\bibitem[Deininger(2011)]{Deininger2011}
Prescott Deininger.
\newblock Alu elements: know the sines.
\newblock \emph{Genome Biology}, 12\penalty0 (12):\penalty0 236, 2011.
\newblock ISSN 1465-6906.
\newblock \doi{10.1186/gb-2011-12-12-236}.
\newblock URL \url{http://dx.doi.org/10.1186/gb-2011-12-12-236}.

\bibitem[Jarvis et~al.(2022)Jarvis, Formenti, Rhie, et~al.]{jarvis2022semi}
E.~D. Jarvis, G.~Formenti, A.~Rhie, et~al.
\newblock Semi-automated assembly of high-quality diploid human reference genomes.
\newblock \emph{Nature}, 611:\penalty0 519--531, 2022.
\newblock \doi{10.1038/s41586-022-05325-5}.
\newblock URL \url{https://doi.org/10.1038/s41586-022-05325-5}.

\bibitem[Ji et~al.(2021)Ji, Zhou, Liu, and Davuluri]{Ji2021}
Yanrong Ji, Zhihan Zhou, Han Liu, and Ramana~V Davuluri.
\newblock Dnabert: pre-trained bidirectional encoder representations from transformers model for dna-language in genome.
\newblock \emph{Bioinformatics}, 37\penalty0 (15):\penalty0 2112–2120, February 2021.
\newblock ISSN 1367-4811.
\newblock \doi{10.1093/bioinformatics/btab083}.
\newblock URL \url{http://dx.doi.org/10.1093/bioinformatics/btab083}.

\bibitem[Lal et~al.(2024)Lal, Garfield, Biancalani, and Eraslan]{Lal2024}
Avantika Lal, David Garfield, Tommaso Biancalani, and Gokcen Eraslan.
\newblock reglm: Designing realistic regulatory dna with autoregressive language models.
\newblock February 2024.
\newblock \doi{10.1101/2024.02.14.580373}.
\newblock URL \url{http://dx.doi.org/10.1101/2024.02.14.580373}.

\bibitem[Liao et~al.(2023)Liao, Zhu, Zhou, Li, Xu, Zhang, and Gao]{Liao2023}
Xingyu Liao, Wufei Zhu, Juexiao Zhou, Haoyang Li, Xiaopeng Xu, Bin Zhang, and Xin Gao.
\newblock Repetitive dna sequence detection and its role in the human genome.
\newblock \emph{Communications Biology}, 6\penalty0 (1), September 2023.
\newblock ISSN 2399-3642.
\newblock \doi{10.1038/s42003-023-05322-y}.
\newblock URL \url{http://dx.doi.org/10.1038/s42003-023-05322-y}.

\bibitem[Liu et~al.(2024)Liu, Li, Hu, Yu, Zheng, Li, Qin, and Yu]{Liu2024}
Junli Liu, Qilin Li, Yixuan Hu, Yi~Yu, Kai Zheng, Dengfeng Li, Lexin Qin, and Xiaochun Yu.
\newblock The complete telomere-to-telomere sequence of a mouse genome.
\newblock \emph{Science}, 386\penalty0 (6726):\penalty0 1141–1146, December 2024.
\newblock ISSN 1095-9203.
\newblock \doi{10.1126/science.adq8191}.
\newblock URL \url{http://dx.doi.org/10.1126/science.adq8191}.

\bibitem[Mo et~al.(2021)Mo, Fu, Hong, Chen, Zheng, Tang, Shen, Xing, and Lan]{GeneBERT}
Shentong Mo, Xi~Fu, Chenyang Hong, Yizhen Chen, Yuxuan Zheng, Xiangru Tang, Zhiqiang Shen, Eric~P Xing, and Yanyan Lan.
\newblock Multi-modal self-supervised pre-training for regulatory genome across cell types, 2021.
\newblock URL \url{https://arxiv.org/abs/2110.05231}.

\bibitem[Nguyen et~al.(2023)Nguyen, Poli, Faizi, Thomas, Birch-Sykes, Wornow, Patel, Rabideau, Massaroli, Bengio, Ermon, Baccus, and Ré]{HyenaDNA}
Eric Nguyen, Michael Poli, Marjan Faizi, Armin Thomas, Callum Birch-Sykes, Michael Wornow, Aman Patel, Clayton Rabideau, Stefano Massaroli, Yoshua Bengio, Stefano Ermon, Stephen~A. Baccus, and Chris Ré.
\newblock Hyenadna: Long-range genomic sequence modeling at single nucleotide resolution, 2023.
\newblock URL \url{https://arxiv.org/abs/2306.15794}.

\bibitem[Nurk et~al.(2022)]{nurk2022complete}
Sergey Nurk et~al.
\newblock The complete sequence of a human genome.
\newblock \emph{Science}, 376\penalty0 (6588):\penalty0 44--53, April 2022.
\newblock \doi{10.1126/science.abj6987}.
\newblock URL \url{https://doi.org/10.1126/science.abj6987}.

\bibitem[Schiff et~al.(2024)Schiff, Kao, Gokaslan, Dao, Gu, and Kuleshov]{Caduceus}
Yair Schiff, Chia-Hsiang Kao, Aaron Gokaslan, Tri Dao, Albert Gu, and Volodymyr Kuleshov.
\newblock Caduceus: Bi-directional equivariant long-range dna sequence modeling, 2024.
\newblock URL \url{https://arxiv.org/abs/2403.03234}.

\bibitem[Schmid \& Deininger(1975)Schmid and Deininger]{Schmid1975}
Carl~W. Schmid and Prescott~L. Deininger.
\newblock Sequence organization of the human genome.
\newblock \emph{Cell}, 6\penalty0 (3):\penalty0 345–358, November 1975.
\newblock ISSN 0092-8674.
\newblock \doi{10.1016/0092-8674(75)90184-1}.
\newblock URL \url{http://dx.doi.org/10.1016/0092-8674(75)90184-1}.

\bibitem[Thakur et~al.(2021)Thakur, Packiaraj, and Henikoff]{thakur2021sequence}
J.~Thakur, J.~Packiaraj, and S.~Henikoff.
\newblock Sequence, chromatin and evolution of satellite dna.
\newblock \emph{International Journal of Molecular Sciences}, 22\penalty0 (9):\penalty0 4309, 2021.
\newblock \doi{10.3390/ijms22094309}.
\newblock URL \url{https://doi.org/10.3390/ijms22094309}.

\bibitem[Vaswani et~al.(2017)Vaswani, Shazeer, Parmar, Uszkoreit, Jones, Gomez, Kaiser, and Polosukhin]{attention}
Ashish Vaswani, Noam Shazeer, Niki Parmar, Jakob Uszkoreit, Llion Jones, Aidan~N. Gomez, Lukasz Kaiser, and Illia Polosukhin.
\newblock Attention is all you need, 2017.
\newblock URL \url{https://arxiv.org/abs/1706.03762}.

\bibitem[Vishniakov et~al.(2024)Vishniakov, Viswanathan, Medvedev, Kanithi, Pimentel, Rajan, and Khan]{vishniakov2024genomic}
Kirill Vishniakov, Karthik Viswanathan, Aleksandr Medvedev, Praveen~K Kanithi, Marco~AF Pimentel, Ronnie Rajan, and Shadab Khan.
\newblock Genomic foundationless models: Pretraining does not promise performance.
\newblock \emph{bioRxiv}, 2024.
\newblock \doi{10.1101/2024.12.18.628606}.

\bibitem[Yang et~al.(2023)Yang, Zhou, Song, et~al.]{yang2023diploid}
C.~Yang, Y.~Zhou, Y.~Song, et~al.
\newblock The complete and fully-phased diploid genome of a male han chinese.
\newblock \emph{Cell Research}, 33:\penalty0 745--761, 2023.
\newblock \doi{10.1038/s41422-023-00849-5}.
\newblock URL \url{https://doi.org/10.1038/s41422-023-00849-5}.

\bibitem[Yoo et~al.(2025)Yoo, Rhie, Hebbar, Antonacci, Logsdon, Solar, Antipov, Pickett, Safonova, Montinaro, Luo, Malukiewicz, Storer, Lin, Sequeira, Mangan, Hickey, Monfort~Anez, Balachandran, Bankevich, Beck, Biddanda, Borchers, Bouffard, Brannan, Brooks, Carbone, Carrel, Chan, Crawford, Diekhans, Engelbrecht, Feschotte, Formenti, Garcia, de~Gennaro, Gilbert, Green, Guarracino, Gupta, Haddad, Han, Harris, Hartley, Harvey, Hiller, Hoekzema, Houck, Jeong, Kamali, Kellis, Kille, Lee, Lee, Lees, Lewis, Li, Loftus, Loh, Loucks, Ma, Mao, Martinez, Masterson, McCoy, McGrath, McKinney, Meyer, Miga, Mohanty, Munson, Pal, Pennell, Pevzner, Porubsky, Potapova, Ringeling, Rocha, Ryder, Sacco, Saha, Sasaki, Schatz, Schork, Shanks, Smeds, Son, Steiner, Sweeten, Tassia, Thibaud-Nissen, Torres-González, Trivedi, Wei, Wertz, Yang, Zhang, Zhang, Zhang, Zhang, Zhao, Zhu, Jarvis, Gerton, Rivas-González, Paten, Szpiech, Huber, Lenz, Konkel, Yi, Canzar, Watson, Sudmant, Molloy, Garrison, Lowe, Ventura, O’Neill, Koren,
  Makova, Phillippy, and Eichler]{Yoo2025}
DongAhn Yoo, Arang Rhie, Prajna Hebbar, Francesca Antonacci, Glennis~A. Logsdon, Steven~J. Solar, Dmitry Antipov, Brandon~D. Pickett, Yana Safonova, Francesco Montinaro, Yanting Luo, Joanna Malukiewicz, Jessica~M. Storer, Jiadong Lin, Abigail~N. Sequeira, Riley~J. Mangan, Glenn Hickey, Graciela Monfort~Anez, Parithi Balachandran, Anton Bankevich, Christine~R. Beck, Arjun Biddanda, Matthew Borchers, Gerard~G. Bouffard, Emry Brannan, Shelise~Y. Brooks, Lucia Carbone, Laura Carrel, Agnes~P. Chan, Juyun Crawford, Mark Diekhans, Eric Engelbrecht, Cedric Feschotte, Giulio Formenti, Gage~H. Garcia, Luciana de~Gennaro, David Gilbert, Richard~E. Green, Andrea Guarracino, Ishaan Gupta, Diana Haddad, Junmin Han, Robert~S. Harris, Gabrielle~A. Hartley, William~T. Harvey, Michael Hiller, Kendra Hoekzema, Marlys~L. Houck, Hyeonsoo Jeong, Kaivan Kamali, Manolis Kellis, Bryce Kille, Chul Lee, Youngho Lee, William Lees, Alexandra~P. Lewis, Qiuhui Li, Mark Loftus, Yong Hwee~Eddie Loh, Hailey Loucks, Jian Ma, Yafei Mao, Juan
  F.~I. Martinez, Patrick Masterson, Rajiv~C. McCoy, Barbara McGrath, Sean McKinney, Britta~S. Meyer, Karen~H. Miga, Saswat~K. Mohanty, Katherine~M. Munson, Karol Pal, Matt Pennell, Pavel~A. Pevzner, David Porubsky, Tamara Potapova, Francisca~R. Ringeling, Joana~L. Rocha, Oliver~A. Ryder, Samuel Sacco, Swati Saha, Takayo Sasaki, Michael~C. Schatz, Nicholas~J. Schork, Cole Shanks, Linnéa Smeds, Dongmin~R. Son, Cynthia Steiner, Alexander~P. Sweeten, Michael~G. Tassia, Fran\c{c}oise Thibaud-Nissen, Edmundo Torres-González, Mihir Trivedi, Wenjie Wei, Julie Wertz, Muyu Yang, Panpan Zhang, Shilong Zhang, Yang Zhang, Zhenmiao Zhang, Sarah~A. Zhao, Yixin Zhu, Erich~D. Jarvis, Jennifer~L. Gerton, Iker Rivas-González, Benedict Paten, Zachary~A. Szpiech, Christian~D. Huber, Tobias~L. Lenz, Miriam~K. Konkel, Soojin~V. Yi, Stefan Canzar, Corey~T. Watson, Peter~H. Sudmant, Erin Molloy, Erik Garrison, Craig~B. Lowe, Mario Ventura, Rachel~J. O’Neill, Sergey Koren, Kateryna~D. Makova, Adam~M. Phillippy, and Evan~E.
  Eichler.
\newblock Complete sequencing of ape genomes.
\newblock \emph{Nature}, April 2025.
\newblock ISSN 1476-4687.
\newblock \doi{10.1038/s41586-025-08816-3}.
\newblock URL \url{http://dx.doi.org/10.1038/s41586-025-08816-3}.

\end{thebibliography}
\bibliographystyle{iclr2025_conference}

\appendix
\section{Appendix}

\begin{figure}[h]
    \centering
    \includegraphics[width=1.0\linewidth]{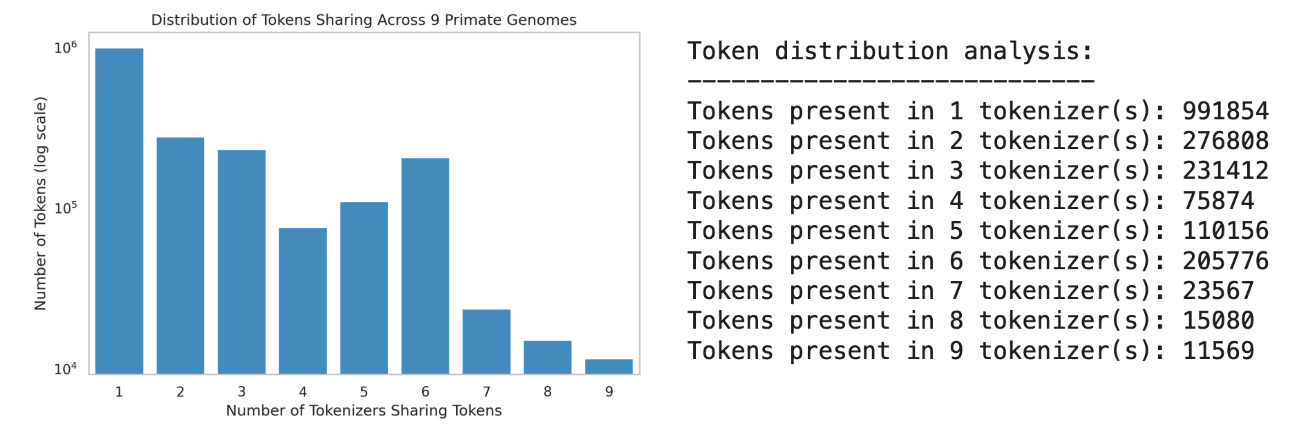}
    \caption{Distribution of token sharing across nine primate genome tokenizers (log scale). The graph shows a dramatic decline in the number of shared tokens as the number of tokenizers increases, from nearly 1 million tokens unique to single genomes to only 11,569 tokens shared across all nine assemblies. The notable drop at four tokenizers reflects the underlying genomic distances: while human assemblies form a tight cluster (Jaccard distances 0.125-0.234), adding any fourth genome introduces substantial vocabulary divergence due to large evolutionary distances (Jaccard distances more than 0.9) between primate species. This pattern demonstrates how species-specific repetitive elements dominate BPE tokenization, challenging the development of a universal genomic tokenizer.}
    \label{fig:supplementary1}
\end{figure}

\end{document}